# Ultracoherent GHz Diamond Spin-Mechanical Lamb Wave Resonators

Xinzhu Li, Ignas Lekavicius[#], Jens Noeckel, and Hailin Wang[*]

Department of Physics, University of Oregon, Eugene, OR 97403, USA

**Abstract**

We report the development of an all-optical approach that excites the fundamental compression mode in a diamond Lamb wave resonator with an optical gradient force and detects the induced vibrations via strain coupling to a silicon vacancy center, specifically, via phonon sidebands in the optical excitation spectrum of the silicon vacancy. Sideband optical interferometry has also been used for the detection of the in-plane mechanical vibrations, for which conventional optical interferometry is not effective. These experiments demonstrate a GHz fundamental compression mode with a $Q$-factor $>10^7$ at temperatures near 7 K, providing a promising platform for reaching the quantum regime of spin mechanics, especially phononic cavity QED of electron spins.

Key words: Nanomechanical resonator, spin-mechanics, silicon vacancy centers, phononic band gap

Mechanical vibrations or phonons play a special role in quantum information platforms. Phonons can mediate coherent interactions between distant solid-state qubits and can provide an interface between solid-state qubits and photons. Major experimental advances have recently been achieved in quantum electromechanics, which couples superconducting qubits to mechanical oscillators. Nonclassical mechanical states including entanglement of mechanical oscillators have been generated[1-3]. Phonon-mediated entanglement between remote superconducting qubits has been demonstrated[4]. A phonon-mediated interface between a superconducting qubit and a photon in the optical domain has also been realized[5]. These advances have stimulated strong and renewed interest in spin-mechanics, which couples spin qubits to phonons in nanomechanical resonators[6,7]. Phonon-mediated coupling between distant electron spins can potentially enable a mechanical quantum network[8-11], providing an experimental platform for spin-based quantum computers. Experimental studies in spin-mechanics have used spin qubits that feature robust spin coherence as well as excellent optical properties, such as negatively charged nitrogen vacancy (NV) and silicon vacancy (SiV) centers in diamond[12,13]. Diamond mechanical resonators such as bulk acoustic wave resonators[14,15], surface acoustic wave (SAW) resonators[16-18], microdisks[19,20], optomechanical crystals[21-23], as well as beams and cantilevers[24-29], have been explored.

Further experimental advance of quantum spin-mechanics requires ultracoherent diamond nanomechanical resonators at a GHz frequency. A mechanical resonator can be characterized by three key parameters: linewidth $\gamma_m$ (or quality factor $Q$), effective mass $m_{eff}$, and resonance frequency $f_m$. Ultrasmall $\gamma_m$ and $m_{eff}$ are essential for reaching the quantum regime of spin-mechanics, while $f_m$ at a GHz frequency is important for reducing effects of thermal phonons and for matching the mechanical resonance with a spin transition. Ultracoherent GHz nanomechanical resonators have been realized with silicon optomechanical crystals embedded in a phononic crystal lattice[30]. A phononic band gap can shield a mechanical oscillator from the surrounding environment and reduce the structural mechanical loss to levels below the intrinsic materials loss[30-35]. Phonon lifetime of a few seconds has been achieved in GHz silicon optomechanical crystals at temperatures of a few mK[30]. Embedding a diamond optomechanical crystal, which integrates a nanomechanical resonator with a photonic crystal optical cavity, in a phononic band gap shield, however, remains difficult[21-23].

A Lamb wave resonator (LWR), which is a thin elastic plate with free boundaries, provides a simple geometry for GHz nanomechanical resonators. As illustrated in Fig. 1a, a rectangular diamond plate with a length of 9.5 μm, embedded in a square phononic crystal lattice, features a fundamental compression mode with $f_m$ near 1 GHz. For the square lattice, the phononic band structure of the symmetric (with respect to the midplane of the plate) compression modes exhibits a large energy gap that protects the compression mode (see Fig. 1b). Figure 1c shows a scanning electron micrograph (SEM) of a diamond phononic structure fabricated with the design shown in Fig. 1a. LWRs can be networked together in a linear or 2D chain in a well-controlled approach[10,36]. This, along with the simplicity of LWRs, the robustness of symmetric compression modes to thickness variations, as well as the protection by a phononic band gap, make these GHz resonators a highly promising platform for spin-mechanics and for the development of mechanical quantum networks of electron spins.

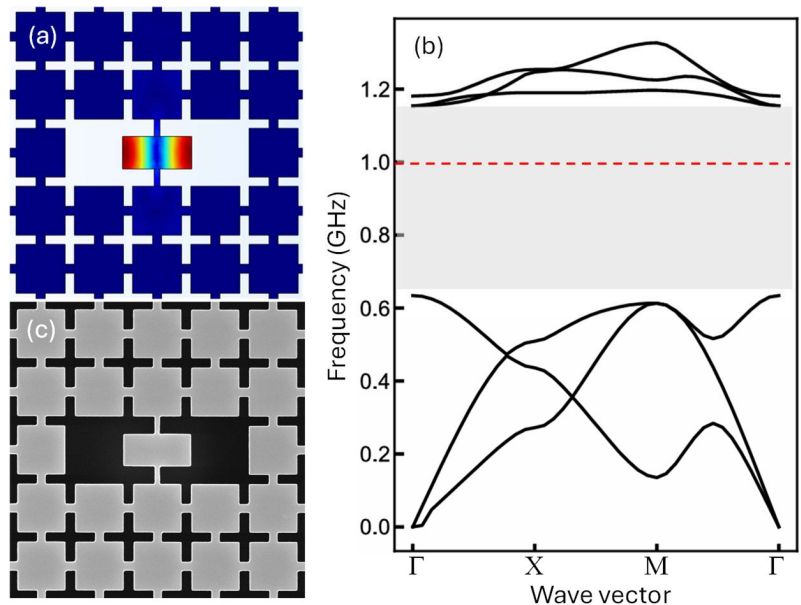

**Fig. 1.** (a) A LWR with dimension (9.5, 4.5) μm embedded in a square phononic crystal lattice with a period of 8 μm, along with the calculated displacement pattern of the fundamental compression mode. The dimension of the bridges in the lattice is (1.25, 1.25) μm. (b) Calculated phononic band structure of the symmetric modes (with respect to the midplane of the plate) of the square lattice with Young's modulus=1200 GPa, Poisson ratio=0.07, and mass density=3500 kg/m³. The phononic band gap shields the fundamental compression mode with a frequency near 1 GHz (the dashed line). (c) SEM of a LWR embedded in a phononic crystal fabricated according to the dimensions given in (a). The sample thickness is 1.5 μm.

The nearly ideal protection and isolation provided by the phononic band gap, however, also make it difficult to excite and detect compression modes in a LWR, since unlike in a cavity-optomechanical system, these modes do not couple to an optical cavity. The conventional approach of using SAWs to excite and detect mechanical vibrations in LWRs is not compatible with achieving ultrahigh $Q$, because by design, the phononic band gap should shield the LWRs from the SAWs. Furthermore, commonly used optical interferometry, while highly sensitive to out-of-plane mechanical displacements, is not effective in detecting in-plane vibrations of the compression modes.

In this paper, we report the development of an all-optical approach to overcome the above obstacles. We have placed a LWR in a focused laser beam and used a temporally modulated optical gradient force to drive the compression mode. The induced mechanical vibrations are probed through their coupling to a SiV center. Resonant excitations of the compression mode by the optical gradient force can induce strong phonon sidebands in the SiV optical excitation spectrum. Mechanical vibrations with an amplitude as small as a picometer can then be detected through sideband optical transitions as well as sideband optical interferometry, instead of the conventional optical interferometry. Using this all-optical approach, we have demonstrated a diamond LWR featuring a fundamental compression mode with $f_m$=0.977 GHz and with $Q$ as high as $1.2 \times 10^7$ at temperatures near 7 K. The $Qf_m$ product achieved is comparable to that of the state-of-the-art silicon optomechanical crystals[30,37] and exceeds that recently reported for diamond optomechanical crystals at mK temperatures[23]. The ultracoherent GHz spin-mechanical resonator and its coupling to spin qubits with excellent optical and spin properties enable a new experimental system for quantum spin-mechanics, especially for phononic cavity QED of electron spins.

The diamond phononic structure shown in Fig. 1c was fabricated from an electronic grade bulk diamond film with $^{28}$Si ions implanted about 45 nm below the diamond surface. A 280 nm layer of $Si_3N_4$ was deposited on the diamond film with plasma-enhanced chemical vapor deposition. This was followed by the deposition of a 500 nm layer of polymethyl methacrylate (PMMA). The etch pattern was defined with electron beam lithography and was transferred from PMMA to $Si_3N_4$ with $CHF_3$ plasma etching. We used $O_2$ plasma reactive ion etching (RIE) with an etching rate of 100 nm/minute to etch the designed pattern on the front side of the diamond film with an etching depth of 1.6 μm. The diamond film was then thinned down from the backside with alternating $Ar/Cl_2$ and $O_2$ plasma RIE until the LWRs were released. The use of a U-shaped shadow mask during the backside etching enabled the released phononic structure to be attached to the bulk diamond film. The processing parameters used were essentially the same as those used in earlier studies[38,39].

The diamond LWR sample was mounted on a cold-finger of a closed cycle cryostat and kept near 7 K. For the resonant excitation of in-plane vibrations, a 1550 nm laser was first sent through an intensity electro-optic modulator (EOM) and then amplified in a fiber amplifier. For photoluminescence excitation (PLE) experiments, a 532 nm laser pulse and a 737 nm laser pulse were used for the initialization and resonant excitation of the SiV center, respectively. All three laser beams were focused onto the sample with a 100x objective (NA=0.85). The 1550 nm laser, with a beam waist (radius) of 2.3 µm, was positioned at the center of the resonator, while the 532 nm and 737 nm lasers were centered at a SiV center. The powers of the lasers were measured after the 100x objective. For sideband optical interferometry, a phase EOM was used for the generation of the two needed optical fields with a fixed relative phase. A schematic of the experimental setup as well as additional information is presented in the supplement[40].

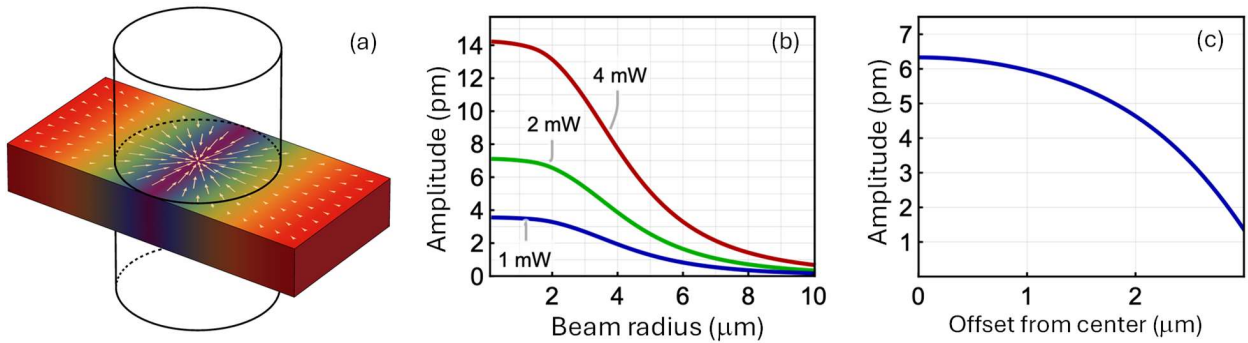

**Fig. 2.** (a) Schematic showing a LWR placed at the waist of a laser beam. The arrows illustrate the directions of the gradient force. (b) Calculated amplitude of the induced mechanical vibration as a function of the laser beam radius, with the incident laser power indicated in the figure. (c) The vibration amplitude decreases as the laser beam, with a power of 2 mW and a radius of 2.25 µm, is offset from the center of the LWR.

As illustrated in Fig. 2a, a diamond LWR with a thickness of $d$ is placed at the waist of a laser beam, normal to the LWR surface. The laser beam exerts two types of forces on the LWR: a radiation pressure force, which scales with the total incident laser power and is ineffective in exciting compression mechanical modes, and a gradient force, which scales with the gradient of the electric field amplitude. At the beam waist, the gradient force points to the center of the laser beam. In the limit that $d \ll$ the depth-of-focus of the laser beam, the areal density of the gradient force in the dipole approximation is given by

$$\mathbf{f}_a(x,y) = \nabla[\mathbf{p}(\mathbf{r}) \cdot \mathbf{E}(\mathbf{r})], \tag{1}$$

where $\mathbf{E}(\mathbf{r})$ is the electric field and $\mathbf{p}(\mathbf{r})$ is the areal density of the induced electric dipoles. For the fundamental compression mode propagating along the x-direction with a mode pattern $\phi(x)$ (see Fig. 1a), we can decompose $\mathbf{f}_a$ into a component $\mathbf{f}_m = F\phi(x)\hat{\mathbf{x}}$, which is matched to $\phi(x)$, and a remainder, which is orthogonal to $\phi(x)$, where $F = \int dx dy \phi(x)\hat{\mathbf{x}} \cdot \mathbf{f}_a(x,y)$ is the mode-matched amplitude and $\hat{\mathbf{x}}$ is the unit vector in the x direction.

For a laser beam with an intensity-modulation frequency, $\omega$, the steady-state mechanical vibration can be written as $u(x,t) = u_\omega \phi(x)\exp(-i\omega t) + c.c.$, with

$$u_\omega = \frac{F}{\rho_a} \frac{1}{\omega_m^2 - \omega^2 - i\omega\gamma_m} \tag{2}$$

where $\rho_a$ is the areal mass density and $\omega_m = 2\pi f_m$. For a theoretical estimate, we consider a LWR with dimension (9.5, 4.5) μm and $\gamma_m/2\pi = 83$ Hz and with corrections to both external and internal optical reflections of the LWR[40]. Figure 2b plots the calculated amplitude of the induced vibration as a function of the laser beam radius. As shown in Fig. 2c, the induced amplitude decreases gradually as the laser beam is offset from the center of the LWR.

The excitation of a longitudinal-acoustic phonon mode in diamond induces a periodic change in the energy gap and a corresponding shift in the energy separation between a ground state, $|g\rangle$, and an excited state, $|e\rangle$ of the SiV center, with the electron-phonon interaction Hamiltonian given by

$$V = i\hbar G(\hat{b} - \hat{b}^+)|e\rangle\langle e|, \tag{3}$$

where $G = D k_m x_{zpf}$ is the electron-phonon coupling rate, with $D$ being the deformation potential and $k_m$ being the wave number of the mechanical mode, $\hat{b}^+$ and $\hat{b}$ are the creation and annihilation operators for the mechanical oscillator, and $x_{zpf} = \sqrt{\hbar/2m_{eff}\omega_m}$ is the zero-point fluctuation. For a dipole optical transition between $|g\rangle$ and $|e\rangle$, the above Hamiltonian leads to sideband optical transitions induced by absorption or emission of the phonons, as illustrated in Fig. 3a. With a laser field at the red sideband of the optical transition, the effective interaction Hamiltonian for the first red sideband transition in the limit of small mechanical amplitude is given by[41],

$$V_r = \frac{\hbar G \Omega_0}{2\omega_m}(\hat{b}|e\rangle\langle g| + \hat{b}^+|g\rangle\langle e|), \tag{4}$$

where $\Omega_0$ is the Rabi frequency for the direct dipole transition, i.e., the carrier transition. The Rabi frequency for the first red sideband transition is thus $\Omega_1 = \Omega_0 G\sqrt{n}/\omega_m$, where $n$ is the average phonon occupation of the mechanical mode. The first blue sideband transition is described by a similar Hamiltonian.

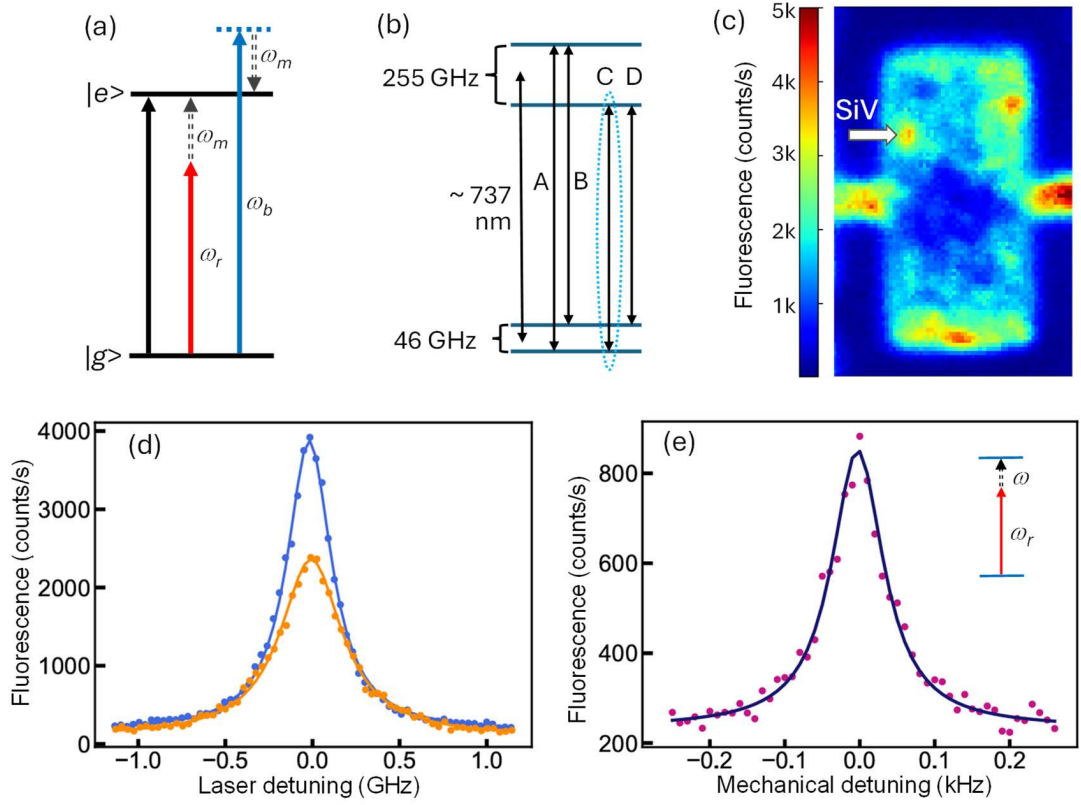

**Fig. 3.** (a) Schematic illustrating the first red and blue sideband transition due to the absorption and emission of a phonon, with a laser frequency, $\omega_r$ and $\omega_b$, near the red and blue sideband of the optical transition, respectively. (b) Energy level structure and optical selection rules for SiV, with the C-transition being highlighted. (c) A confocal fluorescence microscopy image of the LWR indicating the SiV used for the experiment. The image was taken with an excitation laser resonant with the SiV. (d) PLE spectra for the SiV C-transition with (orange dots) and without (blue dots) the application of a CW 1550 nm laser. At zero detuning, the excitation laser is resonant with the C-transition. (e) SiV fluorescence measured as a function of the detuning between the mechanical resonance and the intensity-modulation frequency, $\omega$, of the 1550 nm laser, with $\omega_r$ fixed at nearly 1 GHz below the optical resonance. Solid lines in (d) and (e) are least-square fits to a Lorentzian.

Figure 3b shows the SiV energy level structure and optical selection rules. Experimental studies were carried out on the C-transition ($\lambda$=737.0983 nm) and with a 737 nm laser power of 0.17 μW, unless otherwise specified. The experiments employed a SiV center slightly offset from the center of the resonator, with the SiV position indicated in a confocal fluorescence microscopy image in Fig. 3c. Figure 3d shows the PLE spectrum of the SiV for the C-transition, which features a linewidth of 310 MHz, compared to the radiative-lifetime limited linewidth near 100 MHz. The additional line broadening is in large part due to the power broadening induced by the 737 nm laser[40]. With the application of a 5 mW continuous wave (CW) 1550 nm laser, the PLE linewidth broadens to 460 MHz. Additional studies on line broadening induced by the 737 nm and the 1550 nm laser are discussed in the supplement[40].

In the resolved sideband regime with $\omega_m > \gamma$, where $\gamma$ is the linewidth for the optical transition, the phonon sidebands are spectrally resolved in the SiV excitation spectrum. For the detection of the fundamental compression mode via the first red sideband, the SiV center is subject to an optical field with a frequency fixed at nearly 1 GHz below the optical resonance, while the compression mode is excited by a temporally modulated 1550 nm laser beam with an average power of 2 mW. Figure 3e shows the fluorescence from the SiV center as a function of $\omega$. In this case, the resonant excitation of the compression mode leads to the excitation of the SiV center via the phonon-assisted transition. The mechanical resonance observed in Fig. 3e occurs at $f_m = 0.977$ GHz and features $\gamma_m/2\pi = 83$ Hz, corresponding to $Q = 1.2 \times 10^7$. For comparison, at 10 K silicon optomechanical crystals feature $Q = 5.8 \times 10^5$ with $f_m = 5.1$ GHz [37]. For diamond optomechanical crystals without a phononic crystal shield, a $Q$-factor of $4.4 \times 10^5$ with $f_m$ near 6 GHz has been observed at temperatures near 50 mK[23].

For a detailed study of the phonon sidebands, we measure the PLE spectra of the SiV center, while fixing $\omega$ at the observed mechanical resonance. Figure 4 shows the PLE spectra obtained with increasing average power, $P_m$, for the 1550 nm laser. At relatively small $P_m$, only the first sidebands are observed in the PLE spectra. Higher order phonon sidebands emerge at relatively high $P_m$, indicating the relatively strong excitation of the compression mode by the optical gradient force. As shown in Fig. 4, the amplitude of the first sideband increases initially with $P_m$, corresponding to increased mechanical excitations, and then starts to saturate. This saturation is primarily due to the excitation of second and higher order phonon sidebands. As expected, the increase in the sideband contribution in the PLE spectra is accompanied by a decrease in the amplitude of the carrier resonance, as shown in Fig. 4. The relative amplitudes of the carrier and the sideband transitions can be qualitatively understood without including the line broadening induced by the 1550 nm laser[40]. A quantitative description of the PLE spectra, however, will require a detailed understanding of the effects of 1550 nm laser on the SiV transition.

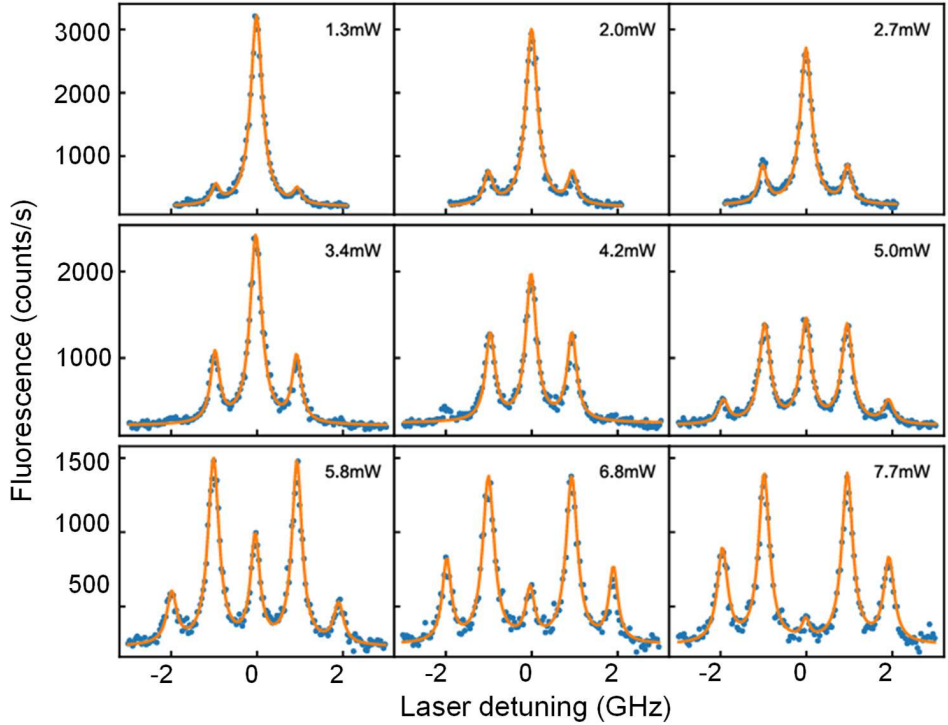

**Fig. 4.** Phonon sidebands in PLE spectra induced by the excitation of the fundamental compression mode. The intensity-modulation frequency for the 1550 nm laser is fixed at the mechanical resonance. The 1550 nm laser power used is indicated in the figure. The solid lines are least-square numerical fits to multiple Lorentzians.

As illustrated in Fig. 5a, both the carrier transition and the red sideband transition can excite the two-level system from $|g\rangle$ to $|e\rangle$. Optical emissions from the excited state thus depend on the relative phase of the two corresponding transition amplitudes. Compression mechanical vibrations can be detected through the interference between these two transition pathways, i.e., through sideband optical interferometry, which should function well even when the spin-mechanical system is not in the resolved sideband regime.

For the experimental implementation, the gradient force generated by an intensity-modulated 1550 nm laser beam, with a modulation frequency $\omega$, excites the fundamental compression mode. The SiV center is driven by two optical fields with frequency $\omega_c$ and $\omega_r$, respectively (see Fig. 5a). These two fields feature a fixed relative phase and a detuning set to $\omega_c - \omega_r = \omega$, for which $\omega_c$ is near the optical resonance. Fluorescence from the SiV center is measured as a function of $\varphi$, the phase of the intensity modulation. The resulting interference fringes shown in Fig. 5b demonstrate the coherent excitation of the compression mode in the LWR. Figure 5c plots the amplitude derived from the respective interference fringes as a function of $\omega$.

The mechanical resonance observed features $f_m$ =0.977 GHz and $\gamma_m/2\pi$ =80 Hz, in good agreement with those obtained from the excitation of the SiV center through the first red sideband shown in Fig. 3e.

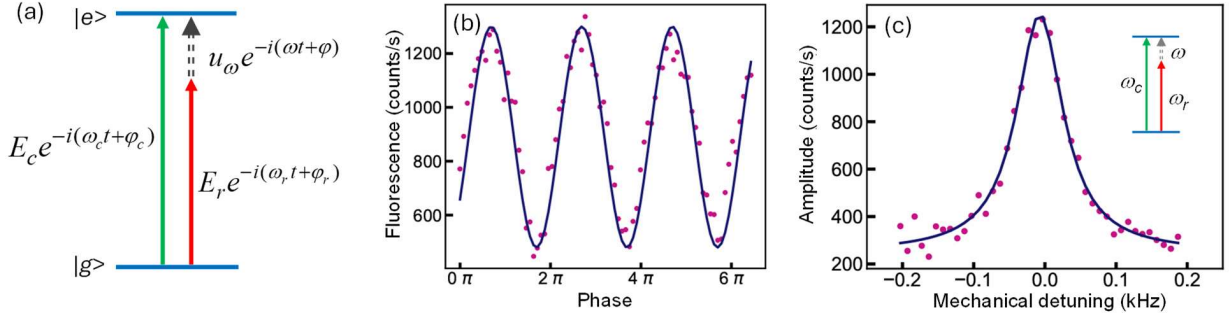

**Fig. 5.** (a) Schematic illustrating the interference between the direct and the first red-sideband transition, where $\omega_c$ is the frequency of the optical field resonant or nearly resonant with the direct transition. (b) Fluorescence from the SiV as function of $\varphi$, the phase of the intensity modulation, with $P_m$=2.7 mW and $\omega/2\pi$= 0.977 GHz. The solid line is a numerical fit to a sinusoidal with a period of $2\pi$. (c) Oscillation amplitudes derived from the respective interference fringes as a function of the detuning from the mechanical resonance. The solid line is a least-square fit to a Lorentzian.

In the limit of weak mechanical excitation, the ratio of the Rabi frequency, $\Omega_1/\Omega_0$, directly measures the amplitude of the mechanical vibration. From Eq. 4, the classical amplitude, $A_m = 2x_{zpf}\sqrt{n}$, of the mechanical vibration is given by,

$$A_m = 2(\Omega_1/\Omega_0)v/D, \qquad (5)$$

for which the SiV is assumed to be at the center of the resonator and $v$ is the acoustic velocity. Without power broadening, $(\Omega_0/\Omega_1)^2$ equals the ratio of the spectrally integrated areas for the carrier resonance and the first red sideband resonance in the corresponding PLE spectrum. Using the PLE spectrum with $P_m$= 2 mW in Fig. 4 and including effects of power broadening in the carrier transition as well as offset of the SiV center from the center of the resonator[40], we estimate that $\Omega_0/\Omega_1$ is near 3. The amplitude of the induced mechanical oscillation is then estimated to be $3\times10^{-12}$ m, for which $v$=1.9x10$^4$ m/s and $D/2\pi$=10$^{15}$ Hz are used[29]. The estimated amplitude is in general agreement with the theoretical expectation shown in Fig. 2b.

Ultracoherent diamond LWRs are especially suitable for phononic cavity QED of electron spins. For a SiV center, direct acoustic transitions between spin states are allowed by the mixing of the spin states induced by an off-axis magnetic field[29]. The single-phonon spin-mechanical
10

coupling rate is thus $g=\eta D k_m x_{zpf}$, where $\eta$ is the effective mixing ratio. The cooperativity for the phononic cavity QED system is $C=4g^2/(\gamma_m \gamma_s)$, where $\gamma_s$ is the linewidth for the spin transition. For the LWR shown in Fig. 1c and with $Q=10^7$, $\eta=0.2$, and $\gamma_s/2\pi=1$ MHz, limited primarily by the $^{13}$C nuclear spin bath[13], $C$ is estimated to exceed 10. Reducing the resonator dimension to (4, 2, 0.3) μm, while keeping other parameters unchanged, leads to $C > 250$. With a further increase in $Q$ as the temperature drops below 100 mK and with $^{12}$C enriched diamond or dynamical decoupling[11], $C>10^6$ can be achieved, for which we take $Q=5\times10^7$ and $\gamma_s/2\pi =1$ kHz. Phononic cavity QED of electron spins with diamond LWRs thus has the potential to rival circuit QED of superconducting qubits.

In conclusion, we have demonstrated the use of optical gradient force to drive compression modes in a diamond LWR. The induced vibrations couple to a SiV center through deformation potential, leading to strong phonon sidebands in the SiV optical excitation spectrum. The vibrations of the compression modes can be effectively detected through the sideband optical transitions as well as sideband optical interferometry. We show that GHz diamond LWRs protected by a phononic band gap can feature a $Q$-factor exceeding $10^7$, enabling a promising platform for quantum spin-mechanics, especially for phononic cavity QED of electron spins. Ultracoherent GHz LWRs can also be extended to other material systems such as silicon carbide that hosts suitable spin qubits[42]. In addition, spin-mechanical LWRs can be networked together, forming a mechanical quantum network of spin qubits and providing a platform for spin-based quantum computers[10,11].

**Supporting information.** Experimental setup, additional experimental results and theoretical analysis.

**Funding sources.** This work is supported by NSF under Grant Nos. 2012524. The earlier fabrication effort was in part supported by the US Air Force Office of Scientific Research.

* Corresponding author email: hailin@uoregon.edu
#Present address: LL: US Naval Research Laboratory, Washington, DC 20375, USA